\documentclass[11pt,a4paper]{article}

\usepackage{amssymb}

\usepackage[utf8]{inputenc}
\usepackage[T1]{fontenc}
\usepackage[english]{babel}
\usepackage[
	top = 0.8cm, includehead,
	bottom = 1cm, includefoot, 
	left = 1.6cm, right = 1.6cm
]{geometry}
\usepackage{indentfirst}

\usepackage{fancyhdr}
\usepackage{caption}
\usepackage{subcaption}
\usepackage{multicol,float}
\usepackage{amsmath,amssymb}
\numberwithin{equation}{section}
\numberwithin{figure}{section}
\makeatletter
\DeclareFontFamily{U}{tipa}{}
\DeclareFontShape{U}{tipa}{m}{n}{<->tipa10}{}
\newcommand{\arc@char}{{\usefont{U}{tipa}{m}{n}\symbol{62}}}%

\newcommand{\arc}[1]{\mathpalette\arc@arc{#1}}
\newcommand{\arc@arc}[2]{%
  \sbox0{$\m@th#1#2$}%
  \vbox{
    \hbox{\resizebox{\wd0}{\height}{\arc@char}}
    \nointerlineskip
    \box0
  }%
}
\makeatother
\usepackage{physics}
\usepackage{siunitx}
\usepackage{xcolor}
\usepackage{graphicx}
\usepackage{tikz}
	\usetikzlibrary{arrows.meta}
\usepackage{fontspec}
\usepackage[default]{fontsetup}
\usepackage[
	backend = biber,
	style = phys,
	biblabel = brackets,
	maxbibnames = 3,
]{biblatex}
\title{\bf
	A New Visual Approach to Pendulum Period Determination
}
\author{
	R. Sánchez-Martínez\textsuperscript{1},\footnote{\ttfamily rodrigo\_smtz@comunidad.unam.mx} \quad
	E. Heredia-Muñoz\textsuperscript{2} \\[2mm]
	\textsuperscript{1}\it\normalsize
	Instituto de Física, Universidad Nacional Autónoma de México. Coyoacán, Cd. de México. México. \\
	\textsuperscript{2}\it\normalsize
	Facultad de Matemáticas, Universidad Veracruzana. Xalapa, Ver. México.
}
\date{\normalsize\today}

\begin{document}
	\maketitle
	\thispagestyle{fancy}
	\begin{abstract}
		The period of oscillation of a simple pendulum ($T = 2\pi\sqrt{l/g}$) is a familiar formula to most first-year physics students. 
    	However, deriving this expression from first principles requires linearizing the equation of motion under the small-angle approximation and solving the resulting differential equation.
    	From our point of view, this method may seem obscure to students in the early stages of learning calculus and lacking in physical insight.
    	Therefore, we propose an alternative approach to the derivation of this formula that relies on geometry, algebra, and physical intuition.
    	Our method follows the foundational idea of integral calculus, replacing the circular path of the pendulum with a successive collection of infinitesimal inclined planes and summing the travel times along each plane as the number of planes becomes very large.
    	Remarkably, evaluating the limit of this sum relies solely on geometric reasoning, making the approach accessible to any student, even those not yet familiar with differential equations or integration techniques.    \end{abstract}
    \hspace*{\fill}\rule{6cm}{0.56pt}\hspace*{\fill}

\normalsize
\begin{multicols}{2}

\section{Introduction and Motivation}
\label{sec:Intro}
	
	The simple pendulum is one of the most extensively studied problems in physics\cite{Yoder1989,Gauld2004,Simonyi2025}.
	As we all know, it consists of a point mass attached to a fixed string, enabling it to swing freely in a back-and-forth motion under the influence of gravity.	
	This simple device captivated the leading figures of the 17th century such as Mersenne, Galileo, Huygens and Newton, and played a crucial role in several key areas: establishing the laws of collision and conservation of energy and momentum, determining the value of the acceleration due to gravity, and, perhaps most importantly, providing evidence for Newtonian synthesis of terrestrial and celestial mechanics \cite{matthews2004ThePendulum}.
	 
	It is well known that, for small initial angular displacements, the oscillating motion of a pendulum can be accurately described by a simple harmonic motion with period
	\begin{equation}
	\label{eq:T_original}
		T = 2\pi\sqrt{\frac{l}{g}}.
	\end{equation}
	\begin{figure*}[ht]
	\centering
	\includegraphics{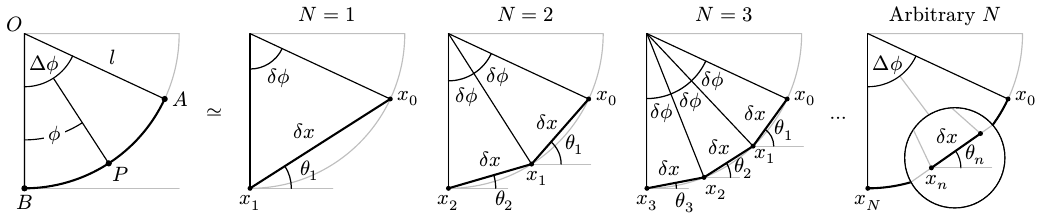}
	\caption{
		Approximation of the circular path followed by the pendulum by a succession of inclined planes.
	}
	\label{fig:pendulum}
	\end{figure*}
	
	This is undoubtedly one of the first formulas learned by the majority of physics students in their introductory year.
	When it is first presented in the classroom, some questions may arise:
	Why doesn't the period depend on the amount of mass attached? Why do longer pendulums swing more slowly than shorter ones? And, most importantly for this work, what is the origin of the $2\pi$ factor in the formula?
	At first glance, one might think it is directly related to the circular motion of the pendulum.
	However, it is evident that the mass does not complete a full circle during an oscillation---which would account for a factor of $2\pi l$ or something similar.
	Instead, it moves along small segments of arcs in a non-uniform manner.
	Hence, if there is indeed a relationship to a circle, it is not the circle traced by the pendulum’s trajectory.
	Theoretical mechanics, in fact, teaches us that the appearance of this factor is due to the periodic relationship of the trigonometric functions used to describe its motion.
	Nonetheless, we would like to emphasize that even without considering any periodicity in the description of the system---as we will demonstrate in this work---a factor of $\pi$ emerges!
		
	Deriving Eq.~\eqref{eq:T_original} from first principles involves writing down the equations of motion of the mass.
	This can be determined using various methods: analyzing the forces acting on the mass and applying Newton's laws, formulating the Lagrangian of the system and solving the Euler-Lagrange equations, or employing the Hamiltonian function and reducing the Hamilton equations.
	Indeed, deriving the equations of motion for a simple pendulum is typically one of the initial examples used in textbooks to illustrate the practical application of a newly introduced classical mechanics formalism (see for example Refs. \cite{taylor2005classical} and \cite{goldstein2011classical}).
	No matter the method employed, one arrives at a second-order non-linear differential equation in terms of the angle whose solution, for arbitrary initial conditions, cannot be expressed in terms of elementary functions.
	Only for sufficiently small initial angular displacements this differential equation can be approximated to that of a simple harmonic oscillator, which has a well-known solution.	
	As a homogeneous linear equation with constant coefficients, it can be solved using the so-called auxiliary equation method,\cite{zill2009} resulting in the familiar periodic solutions in terms of sine and cosine, from which we can extract the expression for the period, Eq.~\eqref{eq:T_original}.
	 While this procedure would serve as an instructive exercise in a basic course on differential equations, for a student who is just learning the basics of differential calculus, this method may seem abstract and lacking in much physical interest.

	The question now is:
	does there exist an alternative procedure to derive expression~\eqref{eq:T_original} that avoids differential equations and integration techniques entirely?
	We claim that there is: following in the footsteps of Archimedes, Newton and Leibniz, we seek to apply the concept of partitioning a complex problem into infinitesimally small, manageable portions to the study of the pendulum’s trajectory.
	Concretely, we replace the circular path of the pendulum with a series of inclined planes, compute the travel time along each using elementary kinematics, and recover the total descent time in the limit as the number of planes grows without bound---much as one evaluates a derivative or integral by returning to its definition rather than invoking established rules.
	In this way, we work directly with the physical and geometric content of the problem, in the spirit of those who first developed the ideas that would later become calculus.

    The structure of this work is as follows:
	First, we describe how the circular path of the pendulum is partitioned into small inclined planes and compute the time required for a particle to traverse each segment using basic kinematic equations.
	Next, we derive an analytical expression for the total falling time as sum of these individual travel times, expressed in terms of the number of planes used.
	The actual falling time for the pendulum is obtained by taking the limit as the number of planes approaches infinity, which we evaluate by means of a visual method based on elementary analytical geometry. 
	Finally, in Section~\ref{sec:Final}, we present our concluding remarks followed by two appendices:
	Appendix~\ref{app:17thCenturyDerivation} provides a historical note on the 17th-century derivation of Eq. \eqref{eq:T_original} and Appendix~\ref{app:AsymptoticEval} presents an asymptotic evaluation of the key limit, establishing its connection to a definite integral for readers already familiar with integral calculus.

	We hope that this work is of great utility and interest to both students and teachers interested in solving mechanics problems using geometric techniques.
	In particular, we encourage teachers to incorporate this derivation into their basic physics classes, inviting them to reproduce our diagrams, since we believe that choosing the appropriate visual elements is an important part of solving a problem and generating meaningful learning.

\section{Our New Derivation}
\label{sec:OurNewDerivation}

	\subsection{Procedure}
	
	Let's examine the motion of a pendulum with length $l$ and mass $m$.
	As shown in Fig.~\ref{fig:pendulum} (first diagram from the left), it starts from rest at point $A$ and falls to point $B$, covering an arc length $\arc{AB} = l\Delta\phi$, where $\Delta\phi$ is the initial angular displacement.	
	At any point $P$ of the trajectory, it is sufficient to know the angle $\phi=\angle BOP$ to determine its position completely.
	Since Eq. (1) is valid for small initial displacements $\Delta\phi \ll 1$, we will work throughout in this regime, whose precise role will become clear in the course of the derivation.
		
	Our strategy is to divide the trajectory $\arc{AB}$ into $N$ inclined planes, each of equal length $\delta x$ and characterized by an inclination angle~$\theta_n$.
	With this approach, we aim to break down the overall motion into manageable segments for analysis.
	Given that the mass experiences a constant acceleration while traversing each plane, it becomes straightforward to formulate an expression for the time interval $\delta_n t$ required for the body to traverse the $n$-th plane.
	Consequently, an approximation to the total falling time from $A$ to $B$ is obtained by summing these intervals, and the precise value emerges as we consider an increasingly large number of inclined planes.
	
\begin{figure}[H]	
	\centering
	\includegraphics{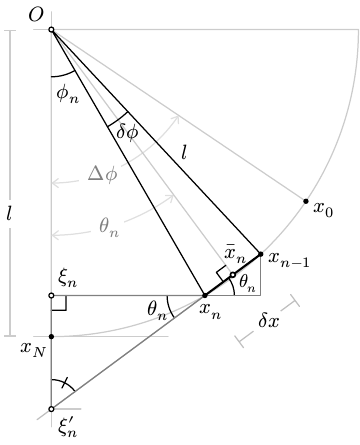}
	\caption{%
		Geometry of the $n$-th inclined plane.
		Note that the triangles $\triangle \xi_n x_n \xi_n'$ and $\triangle \bar{x}_n O \xi_n'$ are similar, since both are right triangles sharing the angle at $\xi_n'$.
		This allows us to write the inclination angle of the $n$-th plane in terms of the angle from the vertical as $\theta_n = \phi_n + \frac{1}{2}\delta\phi$.
	}
	\label{fig:diagram}
	\end{figure}

	The procedure is as follows.
	Let $\{x_0,x_1,\dots,x_N\}$ be a uniform partition of the arc $\arc{AB}$ into $N$ segments of equal arc length, where $x_0 = A$ and $x_N = B$ for a positive integer $N$.
	Since the pendulum moves along a circle of fixed radius $l$, arc length and central angle are proportional---equal arc lengths therefore correspond to equal central angles, ${\delta\phi = \Delta\phi/N}$, and equal chord lengths ${\delta x = 2l\sin(\delta\phi/2)}$.
	With the help of the diagram shown in Fig.~\ref{fig:diagram}, let's analyze the motion of the mass while descending through the $n$-th inclined plane, represented by the line $\overline{x_{n-1}x_n}$ for $n = 1,2,\dots,N$, which has a different inclination angle $\theta_n$ despite subtending the same central angle $\delta\phi$.
	Also, note that the angles to each point $x_n$ from the vertical are ${\phi_n := \angle{x_NOx_n} = (N - n)\delta\phi}$, so that ${\phi_0 = \Delta\phi}$ and ${\phi_N = 0}$.
	
	An expression for the inclination angles of the planes can be found using elementary geometry.
	Let $\overline{x}_n$ be the midpoint between $x_{n-1}$ and $x_n$ and consider the auxiliary points $\xi_n$ and $\xi'_n$, both lying on the vertical from $O$, as shown in Fig.~\ref{fig:diagram}. 
	As described in the caption, the similarity of triangles $\triangle \xi_n x_n \xi_n'$ and $\triangle \bar{x}_n O \xi_n'$ implies that the angles $\angle \xi_n x_n \xi'_n$ and $\angle \xi'_n O \overline{x}_n$ are equal, and therefore,
	\begin{equation}
	\label{eq:theta}
		\theta_n
		= \phi_n + \frac{1}{2}\delta\phi
		= \left(N - \frac{2n - 1}{2}\right)\delta\phi.
	\end{equation}
	
	On the other hand, it is well-known that the acceleration experienced by the mass during its descent along each plane is given by
	\begin{equation}
	\label{eq:a}
		a_n = g\sin\theta_n.
	\end{equation}
	This allows us to calculate the change in speed after each plane using basic kinematic formulas.
	Let $v_n$ be the instantaneous speed acquired after moving past the $n$-th plane, with the initial condition $v_0 = 0$ since the motion starts from rest.
	We can calculate the value of the speed after traversing $n$ planes as follows
	\begin{align}
		\nonumber
		v_0^2 &= 0, \\
		\nonumber
		v_1^2 &= v_0^2 + 2a_1\delta x = 2a_1\delta x, \\
		\nonumber
		v_2^2 &= v_1^2 + 2a_2\delta x = 2(a_1 + a_2)\delta x, \\
		\nonumber
		&\hspace{0.55em}\vdots \\
		\label{eq:vn}
		v_n^2 &= v_{n-1}^2 + 2a_n\delta x = 2(a_1 + \dots + a_n)\delta x.
	\end{align}
	Up to this point, we have kept the initial angular amplitude $\Delta\phi$ unspecified.
	This amplitude plays a crucial role in finding an expression for the speed $v_n$.
	From Eq.~\eqref{eq:vn}, we can see that evaluating $v_n$ requires computing the sum ${\sum_{k=1}^n a_k = g\sum_{k = 1}^n\sin\theta_k}$, which does not admit a simple closed form for arbitrary inclination angles $\theta_k$.
	However, if the initial displacement is sufficiently small, $0< \Delta\phi\ll 1$, then all inclination angles are small as well, and we may replace $\sin\theta_k$ by $\theta_k$ itself.
	To verify this, note from Eq.~\eqref{eq:theta} that the largest inclination angle is $\theta_1 = \big(1-\frac{1}{2N}\big)\Delta\phi \leq \Delta\phi$ for any $N$---so all angles $\theta_k$ are bounded by $\Delta\phi$ and become negligible as $\Delta\phi\to0$.
	As in the traditional derivation of Eq.~\eqref{eq:T_original}, the small-angle regime is what makes the problem tractable---the distinction lies in how each approach exploits it.
	With this approximation, Eq.~\eqref{eq:a} becomes
	\begin{equation}
	\label{eq:a_simplified}
		a_n
		\approx g\theta_n
		= g\left(N - \frac{2n - 1}{2}\right)\delta\phi
	\end{equation}
	and the sum of accelerations can be now evaluated as
	\begin{align}
		\nonumber
		\sum_{k = 1}^n a_k
		= g\sum_{k = 1}^n \theta_k
		& = g\sum_{k = 1}^n \left(N - \frac{2k-1}{2}\right)\delta\phi
		\\
		\nonumber
		& = g\Bigg(Nn - \frac{\sum_{k = 1}^n \big(2k-1\big)}{2}\Bigg)\delta\phi
		\\
		\label{eq:a_sum}
		& = g\left(Nn - \frac{n^2}{2}\right)\delta\phi,
	\end{align}
	where we used the standard identity for the sum of odd numbers $\sum_{k = 1}^n\big(2k - 1\big) = n^2$.
	
	On the other hand, as the number of planes $N$ increases, the length of each plane $\overline{x_{n+1}x_n}$ approaches its corresponding length arc $\arc{x_{n+1}x_n}$, since each chord becomes an increasingly better approximation of the arc it subtends.
	Substituting Eq.~\eqref{eq:a_sum} and $\delta x \approx l\delta\phi$ into the last line of Eq.~\eqref{eq:vn} allows to evaluate
	\begin{align}
		\nonumber
		v_n^2
		= 2\left(\sum_{k = 1}^n a_k\right)\delta x
		&= 2\cdot g\left(Nn - \frac{n^2}{2}\right)\delta\phi \cdot l\delta\phi
		\\
		&= \big(2Nn - n^2\big)gl(\delta\phi)^2.
	\end{align}
	Now, recognizing $2Nn - n^2$ as the difference of two squares, ${N^2 - (N - n)^2}$, we write
	\begin{equation}
		\label{eq:vn_simplified}
		v_n = \sqrt{\big(N^2 - (N - n)^2\big)gl}\;\delta\phi.
	\end{equation}
	
	Finally, from the definition of motion under constant acceleration, we can calculate the time required for the mass to traverse the $n$-th plane using
	\begin{equation}
		\delta_n t = \frac{v_n - v_{n-1}}{a_n}.
	\end{equation}
	In conclusion, the time taken for the actual descent from point $A$ to $B$ is determined by
	\begin{equation}
		t_{AB} = \lim_{N\to\infty}\sum_{n = 1}^N \delta_n t.
	\end{equation}
	Substituting \eqref{eq:a_simplified} and \eqref{eq:vn_simplified} into this expression yields our final result for the total descent time of the pendulum, given in Eq.~\eqref{eq:tAB_large} below.
	\end{multicols}
	
	\noindent
	\tikz[baseline]
		\draw[line width = 0.5pt, {Bar[left,scale = 2.5]}-]
		(0,0) -- (10,0);
	\hspace*{\fill}
	
	\begin{equation}
	\label{eq:tAB_large}
		t_{AB}
		= 2 \lim_{N\to\infty}
		\underbrace{
		\hspace*{-1mm}\left(
			\sum_{n = 1}^N
			\frac{
				\sqrt{N^2 - (N - n)^2} -
				\sqrt{N^2 - (N - (n-1))^2}
			}{
				2N - (2n - 1)
			}
		\right)\hspace*{-1mm}
		}_{F(N)}
		\sqrt{\frac{l}{g}}.
	\end{equation}
	\vspace*{-3mm}
	
	\hspace*{\fill}
	\tikz[baseline]
		\draw[line width = 0.5pt, -{Bar[left,scale = 2.5]}]
		(0,0) -- (10,0);
	
	\begin{multicols}{2}
	\begin{figure*}[ht]
	\centering
	
	\includegraphics{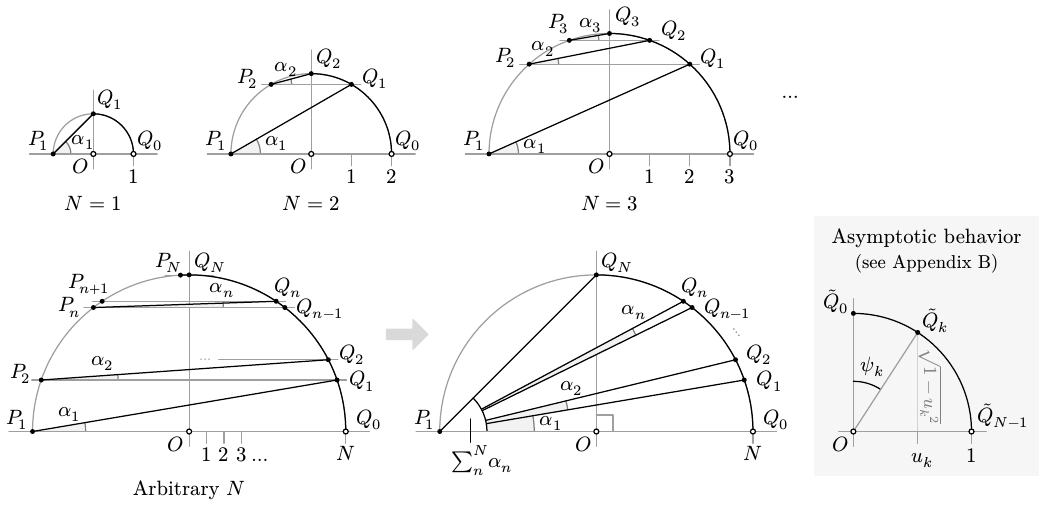}
	\caption{
		Visual aid for the proof of limit \eqref{eq:Pi} and geometrical representation of Eq.~\eqref{eq:sum_of_angles}. 
		Note that the angle $\angle Q_0OQ_N$ is a right angle for any positive integer $N$.
		The inset (bottom right) illustrates the asymptotic behavior as $N\to\infty$: under the reindexing $k = N - n$, the rescaled points $\tilde{Q}_k = Q_k / N$ lie on the unit circle, with horizontal coordinate ${u_k = k/N}$ and arc parameter ${\psi_k = \arccos(u_k)}$.
		See Appendix~\ref{app:AsymptoticEval} for details.
		}
	\label{fig:proof}		
	\end{figure*}

	Hence, the calculation of the period of the pendulum depends on evaluating of the limit of the sum enclosed in parentheses in \eqref{eq:tAB_large}.
	For our result to be consistent with the well-known formula \eqref{eq:T_original}, the following limit must be true:
	\begin{equation}
	\label{eq:Pi}
		\lim_{N\to\infty} F(N) = \frac{\pi}{4}.
	\end{equation}
	We note that, although $F(N)$ is not literally a Riemann sum as written---since each term depends on both $n$ and $N$---, an asymptotic analysis shows that this limit does in fact correspond to the definite integral		
	\begin{equation}
			\int_0^1 \frac{\mathrm{d}u}{2\sqrt{1 - u^2}} = \frac{\pi}{4},
	\end{equation}
		as derived in Appendix~\ref{app:AsymptoticEval}.
	In what follows, however, we establish this result by purely geometric means, making the derivation accessible to students without prior knowledge of integral calculus.
		
	\subsection{Proof of (\ref{eq:Pi})}
	\label{subsec:proof}
	
	Before starting the proof, let's recall that we arrived at Eq.~\eqref{eq:tAB_large} under the only assumption that the pendulum's fall time can be determined by dividing its trajectory into increasingly smaller inclined planes.
	At no point have we introduced the notion that this is a periodic motion ---described by a sine or cosine function---, and therefore, there is no apparent reason for the factor $\pi$ to appear.  
	Yet, it emerges naturally in the proof, serving as a reminder that the resulting equations that describe the motion of a physical system are unique and independent of the formalism applied to derive them.

	While the expression of $F(N)$ in \eqref{eq:tAB_large} may initially appear complex, the limit \eqref{eq:Pi} can be elegantly evaluated through basic analytical geometry and the properties of angles in the circumference.
	With this in mind, we proceed to the proof.
	
	Let $O(0,0)$ be the origin of coordinates in the plane, $N$ a positive integer, and $Q_0(N,0)$ and
	\begin{subequations}
	\begin{align}
		\label{eq:Pn}
		& P_n\left(-(N - n + 1),\sqrt{N^2  - (N - n + 1)^2}\right),
		\\
		\label{eq:Qn}
		& Q_n\left(N - n,\sqrt{N^2  - (N - n)^2}\right),
	\end{align}
	\end{subequations}
	for $n = 1,2,\dots,N$, points on the circumference of radius $N$ and centered at $O$.
	Then, the points $P_n$ and $Q_{n-1}$ are symmetric with respect to the vertical axis, so that $\overline{P_nQ_{n-1}}$ is an horizontal segment.
	The general representation of such points in the circumference is shown in Fig.~\ref{fig:proof}, where the particular cases for $N = 1$ and $N = 2$ are also depicted for illustration.

	From the definition of the points in \eqref{eq:Pn} and \eqref{eq:Qn}, the slope $s_n$ of each segment $\overline{P_nQ_n}$ will be given by
	\begin{equation}
	\label{eq:sn}
		s_n =
		\frac{
			\sqrt{N^2 - (N - n)^2} -
			\sqrt{N^2 - (N - n + 1)^2}
		}{
			(N - n) - \big(-(N - n + 1)\big)
		},
	\end{equation}
	which is precisely the same expression as the terms in the sum of Eq.~\eqref{eq:tAB_large}.
	That is,
	\begin{equation}
		F(N) = \sum_{n = 1}^N s_n.
	\end{equation}

	On the other hand, the slopes $s_n$ can also be written in terms of the respective elevation angle $\alpha_n := \angle Q_{n-1}P_nQ_n$ as $s_n = \tan\alpha_n$.
	Now, given that inscribed angles in a circumference embracing the same arc are equal, we can consider all the inscribed angles $\alpha_n$ sharing a common vertex at $P_1$ (see Fig.~\ref{fig:proof}, rightmost image) to obtain
	\begin{equation}
	\label{eq:sum_of_angles}
		\sum_{n = 1}^N \alpha_n 
		= \sum_{n = 1}^N \angle Q_{n-1}P_1Q_n 
		= \angle Q_0P_1Q_N.
	\end{equation}
		
	But also, we know that any inscribed angle is equal to half of the central angle that spans the same arc, implying that
	\begin{equation}
	\label{eq:pi4}
		\angle Q_0P_1Q_N
		= \frac{1}{2}\angle Q_0OQ_N
		= \frac{\pi}{4}.
	\end{equation}
	It is important to note that this result is independent of the number of points in the partition of the arc.
	That is, it is valid for any positive integer $N$.
	
	This provides an expression for the sum of the angles but we still need an expression for the sum of the slopes, $\sum_{n = 1}^N \tan\alpha_n$, which again cannot be evaluated analytically for arbitrary angles $\alpha_n$.
	However, as the radius $N$ of the circumference tends to infinity (which occurs when considering a very large number of inclined planes), the angles $\alpha_n$ approach zero and we can use again the small-angle approximation ($\tan\alpha_n\approx\alpha_n$) to evaluate the limit \eqref{eq:Pi} as follows,
	\begin{equation}
	\label{eq:final_proof}
		\lim_{N\to\infty} F(N) 
		= \lim_{N\to\infty} \sum_{n = 1}^N \tan\alpha_n 
		= \lim_{N\to\infty} \sum_{n = 1}^N \alpha_n.
	\end{equation}
	But from Eqs.~\eqref{eq:sum_of_angles} and \eqref{eq:pi4}, we have already established that $\sum_{n=1}^N \alpha_n = \pi/4$ regardless of the size of the partition.
	Hence, this result also holds in the limit $N\to\infty$. 
	Consequently, the limit in Eq.~\eqref{eq:final_proof} evaluates to $\pi/4$, completing the proof.

	\section{Conclusion and Final Remarks}
	\label{sec:Final}
	
	Substituting the final result of the proof, Eq. \eqref{eq:final_proof}, into \eqref{eq:tAB_large} provides an expression for the total descent time of the pendulum.
	From this, we can easily calculate the period of one complete oscillation as $T = 4t_{AB}$, thus recovering the well-known expression in Eq.~\eqref{eq:T_original}.
	This concludes our solution to the problem.
	
\begin{figure}[H]	
	\centering
	\includegraphics{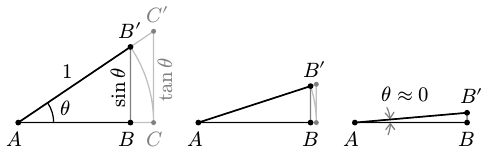}
	\caption{
		Small angle approximation.
		For a right triangle with an arbitrary elevation angle $\theta$ (${0\leq\theta<\pi/2}$), it is evident that the relation $\overline{BB'} < \protect\arc{CB'} < \overline{CC'}$ holds.
		As $\theta$ approaches zero, the points $C$ and $C'$ move increasingly closer to $B$ and $B'$ respectively.
		Consequently, the segments $\overline{BB'} = \sin\theta$ and $\overline{CC'} = \tan\theta$ can be both approximated by the arc length $\protect\arc{CB'} = \theta$. 
	}
	\label{fig:small_angle_approx}
	\end{figure}

	We want to note that, as in the traditional derivation, what makes the problem tractable is the small-angle approximation, employed by our method in two places---first to linearize the acceleration on each inclined plane ($\sin\theta_n \approx \theta_n$), and again to evaluate the key limit ($\tan\alpha_n \approx \alpha_n$).
	Interestingly, it was precisely this realization---that restricting to small oscillations makes the descent time tractable---that led Huygens to derive the formula $T = 2\pi\sqrt{l/g}$ for the first time (see Appendix~\ref{app:17thCenturyDerivation}).
	While a formal derivation of the approximations can be done by means of Taylor series expansion of the trigonometric functions, we believe that in the spirit of this work both uses of the small-angle approximation can be justified geometrically using a diagram such as the one shown in Fig.~\ref{fig:small_angle_approx}.
	Therefore, even students with limited knowledge of the formal machinery of calculus can follow each step and understand the reasoning behind it.
	
	In conclusion, our geometric approach to deriving the pendulum period not only offers an accessible alternative to the traditional derivation---based on differential equations and integration techniques---but also deepens the physical understanding of the problem.
	We hope this method contributes to the teaching and learning of fundamental physics concepts and encourage educators to incorporate similar geometric techniques into their curriculum.
	
	The geometric formulation for solving the problem is attributed to physicist R.~Sánchez-Martínez, while the elegant proof of Eq.~\eqref{eq:Pi} was developed by mathematician E.~Heredia-Muñóz.

\appendix

	\section{17th Century Derivation}
	\label{app:17thCenturyDerivation}
	
	As mentioned in the introduction, no matter what modern method is used to study the pendular trajectory $\phi(t)$, one always arrives at the non-linear differential equation
	\begin{subequations}
	\begin{equation}
		\ddot{\phi} + \frac{g}{l}\sin\phi = 0.
	\end{equation} 
	Then, in the case of $\phi(0) = \Delta\phi \approx 0$, it can be well approximated by the following harmonic oscillator
	\begin{equation}
		\ddot{\phi} + \frac{g}{l}\phi = 0,
	\end{equation}
	\end{subequations}
	with solution $\phi(t) = \Delta\phi \cos(\sqrt{g/l}\, t)$ if the pendulum starts from rest, for example.
	From the periodic properties of the cosine function, we deduce that the amount of time required for the pendulum to complete an oscillation is given exactly by Eq.~\eqref{eq:T_original}. 
	With this in mind, the main idea of this work arose from a simple yet intriguing question: \textit{How could scientists in the 17th century have derived a formula for the period of a pendulum when the concept of differential equations were not yet fully developed?}
	
\begin{figure}[H]
	\centering
	\includegraphics{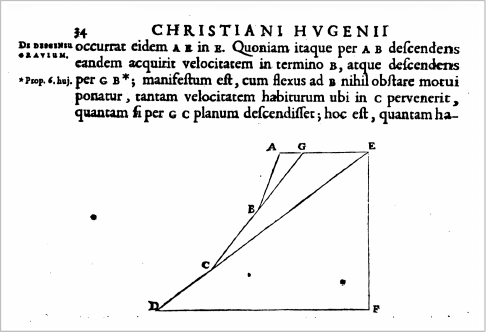}
	\caption{Diagram extracted from Proposition VIII of the second part of Huygen's \textit{Horologium} \cite{huygens1673}.}	
	\label{fig:Huygens}
	\end{figure}

	While searching for an answer, we came across a copy of a Huygens' \textit{Horologium Oscillatorium}, a seminal treatise devoted to the geometric and mechanical study of the pendulum and its application to timekeeping, together with fundamental results on circular motion and what would later be interpreted as centrifugal force.
	In that work, we find the diagram shown in Fig.~\ref{fig:Huygens}.
	Although Huygens used this figure to demonstrate a different principle---that a particle descending through any arbitrary path from a given height reaches the same terminal velocity---this illustration inspired us to explore whether the total descent time of a pendulum could be computed by approximating its trajectory as a succession of inclined planes (compare to Fig.~\ref{fig:pendulum}).
	
\begin{figure}[H]
	\centering
	\includegraphics{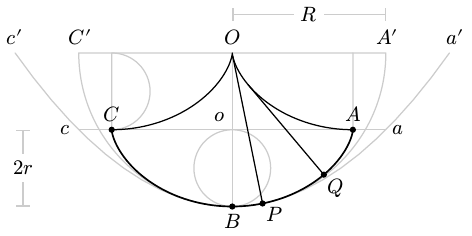}
	\caption{
		Geometric elements of a cycloidal pendulum and the approximation of small oscillations using a parabola.
		Huygens observed that in order to study the descent time of a pendulum from very low positions, such as point $P$, the circular arc $A'BC'$ (of radius $R$) could be approximated by a parabola $a'Bc'$, with focus at $o$--the midpoint of $\overline{OB}$--and latus rectus equal to $2R$. 
		This simplification allowed him to derive, for the first time, a mathematical expression for the period of a simple pendulum, equivalent to Eq.~\eqref{eq:T_original}.
		Clearly, the approximation fails when the pendulum starts from a high position, such as $A'$, where the corresponding point on the parabola, $a'$, differs drastically. 
		Furthermore, Huygens proved that in order for a pendulum to oscillate with the same period regardless of the initial amplitude its path must be a cycloid.
		This can be realized by suspending the pendulum between two cycloidal ``cheeks'' (arcs $\protect\arc{OA}$ and $\protect\arc{OC}$).
		This diagram is a reproduction of \textit{fig.~63} from du Châtelet's \textit{Institutions}\cite{chatelet1740}, with adapted notation to support our explanations.
	}	
	\label{fig:cycloid}
	\end{figure}

	Fortunately, our research also led us to the great work \textit{Unrolling Time: Christiaan Huygens and the mathematization of nature} by Joella G. Yoder~\cite{Yoder1989}.
	In her book, Dr. Yoder reconstructs in great detail the line of reasoning that led Huygens to derive, for the first time, an analytic expression for the descent time of a simple pendulum. 
	In a sense, Huygens followed an idea closely related to the one presented in our work: he sought to compute the total travel time by summing the contributions from successive segments of the circular path. 
	Even without the formal concept of infinitesimals and integration, Huygens was able to evaluate infinite--yet convergent---geometric quantities by cleverly relating their areas to those of other, simpler figures.
	The geometric step-by-step procedure can be somewhat overwhelming to the modern reader and our goal is not to reproduce it here\footnote{See \cite{Yoder1989}, chapter 4.}.
	We merely wish to note that Huygens was likely among the first to recognize that attempting to compute the descent time along an arbitrary circular arc leads to intractable expressions, and that this difficulty motivated him to restrict his analysis to small oscillations.
	In doing so, he discovered that approximating a small circular arc by a parabola was the key to making further progress. 
	This approximation is illustrated in Fig.~\ref{fig:cycloid}.
	
	Having established that his expression for the descent time was valid for sufficiently small oscillations—namely, when the circular arc could be well approximated by a parabola—Huygens went one step further and asked a deeper question: whether there existed a curve for which this property would hold for arbitrary amplitudes.
	This curiosity led him to the discovery of the cycloidal pendulum, whose tautochronous property guarantees that the oscillation period is independent of the initial angle.
	Remarkably, Huygens not only identified the cycloid as the required curve, but also devised a practical method to construct it using fixed cycloidal cheeks, thus translating a purely geometric insight into a realizable mechanical design.

	Finally, although our method differs in scope and purpose, it was largely inspired by our curiosity about how Huygens himself approached the problem of the pendulum period using geometry alone.
	We include this historical note as a way of sharing that curiosity with the reader, and of highlighting how much physical insight can be gained from revisiting classical sources.
	
	\section{Asymptotic Evaluation of $F(N)$}
	\label{app:AsymptoticEval}
	
	In the main text, the limit $\lim_{N\to\infty}F(N) = \pi/4$, Eq.~\eqref{eq:Pi}, was established by means of a purely geometric argument based on the inscribed angle theorem, without appealing to any integration technique.
	However, as noted right after presenting the limit, the same result can be recovered through an asymptotic analysis of the sum $F(N)$, which reveals its connection to a definite integral.
	We present this alternative evaluation here for completeness, and for the benefit of readers already familiar with integral calculus.
	Interestingly, the change of variables required to cast $F(N)$ as a Riemann sum admits a natural geometric interpretation which beautifully aligns to the central-angle analysis used to prove Eq.~\eqref{eq:Pi}.
	We emphasize that this appendix is entirely supplementary---the derivation of the pendulum period in Section~\ref{subsec:proof} is self-contained and independent of the results presented here.  
		
	Recall that $F(N)$ is defined in Eq.~\eqref{eq:tAB_large} as:
	\begin{equation}
		\sum_{n = 1}^N
		\frac
		{\sqrt{N^2-(N-n)^2}-\sqrt{N^2-(N-(n-1))^2}}
		{2N - (2n-1)}.
	\end{equation}
	
	In order to identify this expression as a Riemann sum,
	we begin by substituting $k = N - n$, so that as $n$ runs from $1$ to $N$, $k$ runs from $0$ to $N-1$.
	Using ${N-(n-1) = k+1}$ and ${2N-(2n-1) = 2k+1}$, the sum becomes
	\begin{align}
		\nonumber
		F(N)
		&= \sum_{k=0}^{N-1}\frac{\sqrt{N^2-k^2}-\sqrt{N^2-(k+1)^2}}{2k+1} \\
		&= \sum_{k=0}^{N-1}\frac{N\left(\sqrt{1-\big(\frac{k}{N}\big)^2}-\sqrt{1-\big(\frac{k+1}{N}\big)^2}\right)}{2k+1}.
	\end{align} 
	Now, since $k$ always appears divided by $N$, we introduce the rescaled variable ${u_k = k/N}$ with ${\delta u = u_{k+1} - u_k = 1/N}$, so $u_k$ takes values in $[0,1)$, and $\delta u \to 0$ as long as $N\to\infty$.
	The denominator becomes
	\begin{equation}
		2k + 1 = 2Nu_k + 1 \approx 2Nu_k = \frac{2u_k}{\delta u},
	\end{equation}
	where the approximation holds for large $N$ and $k\geq1$.
	This gives\footnote{%
		The approximation of the denominator breaks down at ${k = 0}$ since both numerator and denominator vanish. 
		However, this term can be evaluated directly,
		\begin{equation*}
			\frac{\sqrt{N^2 -0}-\sqrt{N^2-1}}{1}
			= N - \sqrt{N^2-1}
			= \frac{1}{N+\sqrt{N^2-1}}
			\overset{\strut N\to\infty}{\longrightarrow} 0,
		\end{equation*} 
		so it contributes nothing to the limit and the Riemann sum argument holds for all remaining terms.
	}
	\begin{equation}
		F(N)
		= \sum_{k = 0}^{N-1}\frac{N\left(\sqrt{1-u_k^2} - \sqrt{1-(u_k+\delta u)^2}\right)}{2u_k/\delta u}
	\end{equation}
	
	The difference in the numerator can be regarded as a finite difference of the function ${f(u_k) = \sqrt{1 - u_k^2}}$ and can be evaluated as follows.
	From differential calculus we know that
	\begin{equation}
		\lim_{\delta u\to0}
		\frac{f(u_k + \delta u) - f(u_k)}{\delta u}
		= f'(u_k).
	\end{equation} 
	Therefore, for sufficiently small $\delta u$, the approximation ${f(u_k + \delta u) - f(u_k) \approx f'(u_k)\,\delta u}$ is valid and, in particular,
	\begin{equation}
		\sqrt{1 - u_k^2} - \sqrt{1 - \big(u_k + \delta u\big)^2} \approx \frac{u_k}{\sqrt{1 - u_k^2}}\cdot\delta u.
	\end{equation}
	Substituting this result into $F(N)$, the factors of $N$, $u_k$, and $\delta u$ simplify as follows
	\begin{equation}
		F(N)
		\approx \sum_{k = 0}^{N-1}\frac{N\cdot\cfrac{u_k}{\sqrt{1-u_k^2}}\cdot\delta u}{\cfrac{2u_k}{\delta u}} 
		= \sum_{k = 0}^{N-1}\frac{\delta u}{2\sqrt{1 - u_k^2}}.
	\end{equation}
	Now, it is straightforward to promote the discrete sum to a continuous integral as
	\begin{equation}
	\label{eq:intF}
		\lim_{N\to\infty} F(N)
		= \int_{0}^1\frac{\mathrm{d}u}{2\sqrt{1-u^2}},	
	\end{equation}
	which can be evaluated exactly as $\frac{1}{2}\arcsin(u)\big|_{0}^{1} = \pi/4$.
	
	Moreover, the substitution $u_k = k/N$ introduced above admits a natural geometric interpretation in terms of the points $Q_n$ defined in the Section~\ref{subsec:proof}.
	Recall that those points lie on a circle of radius $N$ centered at the origin, with coordinates given by Eq.~\eqref{eq:Qn}.
	Rewriting in terms of $k = N - n$, these are ${Q_k = (k, \sqrt{N^2 - k^2})}$.
	Factoring $N$, the rescaled points $\tilde{Q}_k = Q_k / N$ have coordinates
	\begin{equation}
		\tilde{Q}_k
		= \Big(k/N,\sqrt{1 - (k/N)^2}\Big)
		= \Big(u_k, \sqrt{1 - {u_k}^2}\Big),
	\end{equation}
	which lie on the unit circle.
	Writing ${u_k = \cos\psi_k}$, each rescaled point takes the familiar form
	\begin{equation}
		\tilde{Q}_k = (\cos\psi_k, \sin\psi_k),
	\end{equation}
	where the arc parameter $\psi_k = \arccos(u_k)$ runs from ${\psi = \pi/2}$ at ${k = 0}$ to $\psi \approx 0$ at ${k = N-1}$.
	Therefore, $u_k$ is simply the horizontal coordinate of the rescaled point $\tilde{Q}_k$ on the unit circle, or equivalently, the cosine of the angle $\psi_k$ subtended at the origin.
	Note that this direction is opposite to that of the original indexing in $n$, as illustrated in the inset of Fig.~\ref{fig:proof}. 
	As $N\to\infty$, the discrete set of points $\tilde{Q}_k$ densely fills the quarter arc of the unit circle from $(0,1)$ to $(1,0)$, and the variable $u\in[0,1)$ parametrizes this arc continuously.
	
	In this light, the integral in Eq.~\eqref{eq:intF} acquires a transparent geometric meaning.
	Indeed, substituting $u = \cos\psi$, so that $\mathrm{d}u = -\sin\psi\,\mathrm{d}\psi$ and the limits $u\in[0,1]$ correspond to $\psi\in[0,\pi/2]$, the integral reduces to:
	\begin{equation}
		\int_{0}^1\frac{\mathrm{d}u}{2\sqrt{1-u^2}}
		= \int_{0}^{\pi/2}\frac{\mathrm{d}\psi}{2} 
		= \frac{\pi}{4}.
	\end{equation}
	The integral thus sweeps the quarter arc of the unit circle, and its value $\pi/4$ is nothing other than half the total angle subtended by that arc---in perfect agreement with the inscribed angle ${\angle Q_0P_1Q_N = \pi/4}$ established in Eq.~\eqref{eq:pi4}, revealing a deep consistency between the geometric proof of Section~\ref{subsec:proof} and the asymptotic analysis presented here.
	
\printbibliography
\end{multicols}
	
\end{document}